\documentclass[iop]{emulateapj}
\usepackage{textcomp}
\usepackage{graphicx}
\usepackage{color}

\shorttitle{New constraints on the nature of the martian moons' building blocks}

\shortauthors{Ronnet et al.}

\begin{document}

\title{Reconciling the orbital and physical properties of the martian moons}

\author{T. Ronnet\altaffilmark{1}, P. Vernazza\altaffilmark{1}, O. Mousis\altaffilmark{1}, B. Brugger\altaffilmark{1}, P. Beck\altaffilmark{2}, B. Devouard\altaffilmark{3}, O. Witasse\altaffilmark{4}, F. Cipriani\altaffilmark{4}}

\affil{$^1 \,$Aix Marseille Universit{\'e}, CNRS, LAM (Laboratoire d'Astrophysique de Marseille) UMR 7326, 13388, Marseille, France}
\affil{$^2\,$Univ. Grenoble Alpes, IPAG, F-38000 Grenoble, France}
\affil{$^3\,$Aix-Marseille Universit{\'e}, CNRS, IRD, CEREGE UM34, 13545 Aix en Provence, France}
\affil{$^4\,$European Space Agency, ESTEC, Keplerlaan 1, 2200 AG Noordwijk, The Netherlands}

\begin{abstract}

The origin of Phobos and Deimos is still an open question. Currently, none of the three proposed scenarios for their origin (intact capture of two distinct outer solar system small bodies, co-accretion with Mars, and accretion within an impact-generated disk) is able to reconcile their orbital and physical properties. Here, we investigate the expected mineralogical composition and size of the grains from which the moons once accreted assuming they formed within an impact-generated accretion disk. A  comparison of our results with the present day spectral properties of the moons allows us to conclude that their building blocks cannot originate from a magma phase, thus preventing their formation in the innermost part of the disk. Instead, gas-to-solid condensation of the building blocks in the outer part of an extended gaseous disk is found as a possible formation mechanism as it does allow reproducing both the spectral and physical properties of the moons. Such a scenario may finally reconcile their orbital and physical properties alleviating the need to invoke an unlikely capture scenario to explain their physical properties.

\end{abstract}


\section{Introduction}

During the 70s and 80s, dynamicists have demonstrated that the present low eccentricity, low inclinations and prograde orbits of Phobos and Deimos are very unlikely to have been produced following capture  \citep{bur78,pol79}, thus favoring a formation of the moons around Mars \citep{sze83,caz80,gol63}. Despite such early robust evidence against a capture scenario, the fact that the moons share similar physical properties (low albedo, red and featureless VNIR reflectance, low density) with outer main belt D-type asteroids has maintained the capture scenario alive \citep{fra12,fra14,paj13}. 

\begin{figure*}[t]
\begin{center}
\includegraphics[scale=0.6]{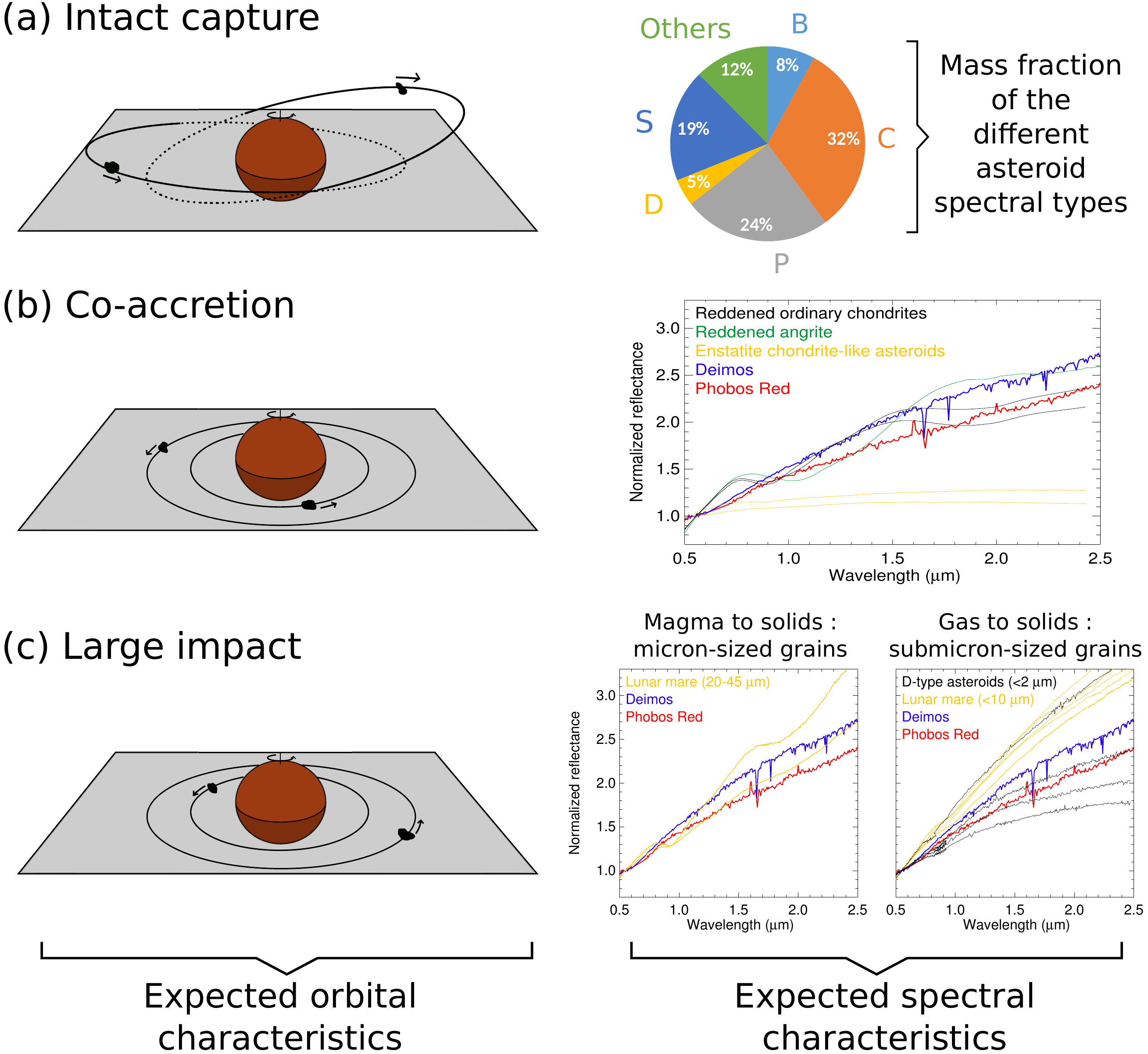}  
\caption{Schematic representation of the expected orbital (left) and spectral (right) characteristics of the martian moons for each of the three different scenarios currently invoked for their origin. Note that in the case of Phobos, we display the average spectrum of the red region. The Phobos and Deimos spectra are CRISM/MRO data that were retrieved from PDS: http://pds-geosciences.wustl.edu/. The lunar mare spectra were retrieved from: http://pgi.utk.edu/. The meteorite spectra were retrieved from RELAB: www.planetary.brown.edu/relab/. The asteroid spectra were retrieved from: http://smass.mit.edu/. (a) The intact capture scenario would likely produce retrograde, large, eccentric and inclined orbits. The nearby asteroid belt being a good proxy for the asteroid types that could have been captured, we display the spectral diversity of the latter \citep{dem13}. Both D-type asteroids (which are the closest spectral analogs to Phobos and Deimos) and P-types are thought to have formed in the primordial trans-Neptunian disk and to have been injected in the inner solar system during the late migration of the giant planets (e.g., the Nice model; \citep{lev09}). Such event could have potentially led to a few of these objects being captured as moons by Mars. The problem with this scenario is that P-types are twice as abundant as D-types;  the capture of two D-types around Mars rather than two P-types or even one P-type and one D-type is thus not statistically favored. Along these lines, an additional caveat of the capture scenario is that the density of the largest (D $\geq$ 200km) P- and D-type asteroids lies in the 0.8-1.5 g/cm$^{3}$ range \citep{Car012}. Density decreasing with asteroid size for a given composition \citep{Car012}, we would expect the density of Phobos and Deimos to be somewhere in between the density of the comet 67P ($\sim$0.5 g/cm$^{3}$; \citep{Sie015}) and the one of the largest P and D-types \citep{Car012}, thus clearly below the one of the Martian moons.
(b) In the co-accretion scenario, circular and co-planar orbits would be expected and the spectral characteristics of the martian moons would likely ressemble those of either reddened ordinary chondrites, reddened angrites or enstatite chondrite-like asteroids (note: enstatite chondrites barely redden via space weathering effects - see \citet{ver09}). Yet, this is not the case. (c) Within the impact scenario, a condensation directly from a magma (left) would lead to the martian moons having typical lunar mare like spectral properties resulting from the coexistence of fine ($\leq$10 microns; spectrally featureless) and of large ($\geq$10 microns; spectrally feature-rich) olivine and pyroxene grains at their surfaces. Alternatively, gas-to-solid condensation in the external part of the disk (right) would lead to the formation of small grains ($\leq$2 microns) and thus naturally explain the similarity in spectral properties between the moons and both D-type asteroids and fine grained ($\leq$10 microns) lunar soils.} \label{fig1}
\end{center}  
\end{figure*}

Whereas the present orbits of the moons are hardly compatible with a capture scenario, they correspond to the expected outcome of an in situ formation scenario either as the result of co-accretion or of a large impact. Co-accretion with Mars appears unlikely because Phobos and Deimos would consist of the same building block materials from which Mars once accreted. Those building blocks would most likely comprise water-poor chondritic meteorites (enstatite chondrites, ordinary chondrites) and/or achondrites (e.g., angrites), which are all suspected to have formed in the inner ($\leq$2.5 AU) solar system, namely interior to the snowline. This assumption is supported by the fact that the bulk composition of Mars can be well reproduced assuming ordinary chondrites (OCs), enstatie chondrites and/or angrites as the main building blocks \citep{san99, bur04,fit16}. Yet, OCs as well as the remaining candidate building blocks (enstatite chondrites, angrites) are spectrally incompatible with the martian moons, even if space weathering effects are taken into account (see panel b in Figure~\ref{fig1}).  

It thus appears from above that accretion from an impact-generated accretion disk remains as the only plausible mechanism at the origin of the martian moons. As a matter of fact, the large impact theory has received growing attention in recent years \citep{cra11,ros12,can14,cit15}. This hypothesis is attractive because it naturally explains the orbital parameters of the satellites as well as some features observed on Mars such as (i) its excess of prograde angular momentum possibly caused by a large impact \citep{cra11}, and (ii) the existence of a large population of oblique impact craters at its surface that may record the slow orbital decay of ancient moonlets formed from the impact-generated accretion disk \citep{sch82}. Along these lines, \citet{cit15} have recently shown that a large impact (impactor with 0.01--0.02 Mars masses) would generate a circum-Mars debris disk comprising $\sim$1--4\% of the impactor mass, thus containing enough mass to form both Phobos and Deimos. Although the impact scenario has become really attractive, it has not yet been demonstrated that it can explain the physical properties and spectral characteristics of the martian moons. 

Here, we investigate the mineralogical composition and texture of the dust that would have crystallized in an impact-generated accretion disk. Since there are no firm constraints regarding the thermodynamic properties of the  disk, we perform our investigation for various thermodynamic conditions and impactor compositions. We show that under specific disk's pressure and temperature conditions, Phobos and Deimos' physical and orbital properties can be finally reconciled.

\section{Formation from a cooling magma}

Because of the absence of constraints regarding the composition (Mars-dominated or impactor-dominated) and the thermodynamic conditions of the impact-generated disk, several configurations must be investigated in order to understand the formation conditions of Phobos and Deimos within such a scenario. As a first step, we considered the protolunar disk as a reference case because it is so far the most studied impact-generated accretion disk. Its structure has been investigated by \citet{tho88} and subsequently by \citet{war12,war14}. These studies have shown that the disk's midplane consists of a liquid phase surrounded by a vapor atmosphere. Beyond the Roche limit, gravitational instabilities developed and large clumps formed directly from the magma \citep[e.g.][]{sal12,kok00}. Those clump then agglomerated to form the Moon. In this case, because of internal evolutionary processes (differentiation, convection, etc..), the mineralogical composition and thus the spectral properties of the lunar mantle and crust will differ from the clump' ones. In the martian case, the situation is different in the sense that the clumps possess right away a mass/size comparable to that of Phobos and Deimos \citep{ros12}. This implies that the final composition and spectral properties of the martian moons would directly reflect those of the minerals that crystallized from the magma disk. 

\subsection{Methods}

We considered three different compositions for the impactor (see Table~\ref{tbl1}), namely a Mars-like composition (1), a Moon-like composition (2) and an outer solar system composition (3) (i.e., TNO). The latter case would be coherent with an inward migration of a large planetesimal as a consequence of the possible late migration of the giant planets \citep[e.g.,][]{mor05}. By analogy with the Earth-Moon system, it has been suggested, however, that the impactor most probably formed near the proto-Mars \citep[][see cases (1) and (2)]{har75} but one cannot exclude that the impactor formed elsewhere (3). 

In addition, since the relative proportions of the impactor and martian materials are poorly constrained in the resulting disk, we considered various proportions between these two materials. We considered two cases, namely a disk exclusively made of the impactor mantle and a half--half fraction. Case (1) complements this sequence by illustrating the case for a 100\% fraction of the martian mantle.


To estimate the composition of the solids crystallized from the magma and thus of the moons, we performed a CIPW normative mineralogy calculation \citep{Gon016}. 
This method allows determining the nature of the most abundant minerals that crystallize from an anhydrous melt at low pressure while providing at the same time a  good estimate of their final proportions.
The CIPW norm calculation is well adapted to our case given that the disk supposedly cooled down slowly through radiation \citep{war12}, allowing complete crystallization of the minerals. It should be noted that we do not aim at determining the exact composition of the moons. Considering the few constraints we have on the system, our purpose is to discriminate between plausible and unplausible scenarios and thus provide new constraints for future studies.

\subsection{Results}

In this section, we present the inferred mineralogical composition of the moons for the three aforementioned impactor compositions (see Table~\ref{tbl2}) and for the different relative abundances of the impactor and martian mantle.

\begin{itemize}

\item\textit{Case 1 (Mars-like impactor)} : Since the impactor has a composition similar to that of Mars, we performed the calculation using a Bulk Silicate Mars (BSM) magma composition \citep[taken from][]{lod11}. The BSM is an estimate of the chemical composition of Mars' mantle. By calculating the CIPW norm, we found that both olivine and orthopyroxene (hypersthene) are the main minerals to crystallize ($\sim$59\% and $\sim$21\%  respectively). Both diopside and feldspar (plagioclase) are also formed although in significantly lower proportions ($\sim$7\% and $\sim$12\%  respectively). 

Note, however, that the above results do not account for a partial vaporization of the disk. The fraction of vaporized material is speculative although theoretical considerations advocate that it should be more than 10\% in the case of the protolunar disk \citep{war12,war14}. To emphasize the role of vaporization on the resulting composition of the building blocks of the moons, we considered the case of a half vaporized disk (see Table~\ref{tbl1}b). Its magma composition was derived following the results of \citet{can15} for a Bulk Silicate Earth (BSE) disk's composition. This first order approximation is quantitatively valid as the BSE and a BSM compositions are very similar \citet{vis13}. By applying the CIPW norm to this new magma composition, we found that significantly more olivine is crystallized ($\sim$85\%), whereas both orthopyroxene and diopside do not form. The proportion of feldspar remains, however, the same ($\sim$10\%).


\item\textit{Case 2 (Moon-like impactor)} : Here, we used the Bulk Silicate Moon composition as a proxy for the impactor composition. For both a 50-50\% Moon-Mars mixing ratio and a pure lunar-like composition, we found that both olivine and orthopyroxene (hypersthene) are the main crystallizing minerals ($\sim$60\% and $\sim$22\%  respectively). In both cases, it thus appears that the derived bulk composition of the moons is very close to the one obtained for a Bulk Silicate Mars disk's composition. Taking into account a partial vaporization of the magma would also lead to results similar to those obtained for \textit{case 1}.

\item\textit{Case 3 (TNO-like impactor)} : Here we used the composition of interplanetary dust particles \citep[IDPs;][]{rie09} as a proxy for the composition of the TNO-like impactor. IDPs which are the likely building blocks of comets may also be the ones of TNOs if one follows the basic and currently accepted assumption that both population formed in the outer solar system. However, by using directly the composition of IDP grains, we neglect the effect of differentiation that has likely occurred on a Moon-sized TNO. This implies that we certainly overestimate the amount of iron in the disk. 

When considering a pure IDP-like composition, quartz crystallizes because of an excess of silica. Indeed, the amount of Mg and Fe does not allow to form enough olivine and pyroxene to account for all the available Si. Moreover, quartz and Mg-rich olivine being mutually exclusive minerals, the absence of one of the two is the norm if the other is formed. In this case, the resulting composition is pyroxene-rich instead of olivine-rich. A substantial amount of pyrite is also formed due to the high proportion of sulfur in IDPs. When considering a 50-50\% TNO-Mars mixing ratio, there is no longer an excess of silica. Orthopyroxene remains the most abundant mineral but olivine is formed instead of quartz and in a larger amount.

\end{itemize}

In summary, we find that for every tested scenario the inferred mineralogical composition of the building blocks of the moons (and thus of the moons) is either olivine-rich or pyroxene-rich. Since minerals that solidify from a slow cooling magma are usually coarse grained \citep[grain size usually in the 10$\mu$m-1mm range; see A1;][]{cas93, sol07}, our findings imply that if Phobos and Deimos actually formed from a disk of magma, then their spectra --similarly to those of either S-type asteroids or lunar mares-- should display detectable 1 and 2 micron bands (see Fig. 1, panel c - left)  that are characteristic of the presence of olivine (1 micron) and pyroxene (1 and 2 microns). 

Yet, this is not the case. It is very unlikely that space weathering effects --which are more significant at 1 AU than at 1.5 AU-- could suppress the olivine and pyroxene absorption bands in the martian moons spectra considering that those effects are not able to suppress them in the lunar ones \citep{pie00,yam12}. We thus conclude that it is highly unlikely that Phobos and Deimos actually formed from a disk of magma. Another argument in disfavor of this scenario is given by the fact that the magma resides inside the Roche limit (at $\sim$4$R_\mathrm{Mars}$) which in the martian case is located inside the synchronous orbit (at $\sim$6$R_\mathrm{Mars}$). Thus, the bodies that formed directly from the magma must have impacted Mars a long time ago as a consequence of their orbital decay due to tidal forces \citep{ros12}.  

\section{Formation in an extended gaseous disk}

A different formation mechanism is thus required to explain both the current orbits of the moons as well as their spectral characteristics. Of great interest, \citet{ros12} suggested that the moons could have formed in an extended gaseous disk farther from Mars than in the two-phase disk case. Such a disk should be initially hot so that it would thermally expand under pressure gradients and be gravitationally stable beyond the Roche limit.
 As the disk would thermally expand, it would cool down rapidly. The extended disk would also have a lower pressure and a larger radiative surface allowing, again, a faster cooling than a compact disk residing inside the Roche limit.

In this scenario, the conditions under which Phobos and Deimos formed could have been similar to those that occurred in the protosolar nebula in the sense that small solid grains could have condensed directly from the gas without passing through a liquid (magma) phase (see A2). In this case, Phobos and Deimos would consist of material that has the same texture --not necessarily the same composition-- as the one that has been incorporated into comets and D-type asteroids, namely fine grained dust (grain size $\leq$ 2 microns; see A2; \citet{ver15}). It would therefore not be surprising that these objects share similar physical properties, including i) a low albedo (pv$\sim$0.06) and ii) featureless and red spectral properties in the visible and near-infrared range (see Fig.~\ref{fig1} panel c -right, \citet{ver15}). It is important to stress here that there are currently no available laboratory reflectance spectra in the visible and near-infrared range for sub-micron sized particles. The spectral behavior in this wavelength range can be reproduced via the Mie theory as implemented by \citet{ver15}. These authors showed that a space weathered mixture of sub-micron sized olivine and pyroxene grains would possess spectral properties similar to those of P- and D-type asteroids. Future laboratory measurements will be necessary in order to characterize the reflectance properties of all kinds of minerals (silicates, phyllosilicates, iron oxides, etc..) in order to provide more accurate constraints on the composition of the moons. Note that a further comparison of the Phobos and Deimos spectra with those of fine grained lunar mare soils (grain size $\leq$ 10 microns) reinforces the idea that the moons may effectively be aggregates of sub-micron sized grains. Indeed, although the average grain size of the lunar soils is small ($\leq$ 10 microns), absorption bands at 1 micron are still visible, suggesting that the grains at the surface of the martian moons must be even smaller than these already fine grained lunar samples.

Finally, accretion from such poorly consolidated sub-micron sized material would also naturally explain the low densities ($\sim$1.86 g/cm$^{3}$ for Phobos and $\sim$1.48 g/cm$^{3}$ for Deimos) and high internal porosities ($\sim$40-50\% assuming an anhydrous silicate composition) of the moons \citep{and010,ros11,wil14}. Importantly, such high porosity is not observed in the case of S-type asteroids with diameters in the 10-30 km size range (a $\sim$20-30\% porosity is observed for these objects; \citet{Car012}), reinforcing the idea that the building blocks of the martian moons must be drastically different in texture from those of S-type asteroids (i.e., OCs). 

 
The fact that the building blocks of the moons should avoid a magma phase provides interesting constraints on the thermodynamic conditions that prevailed inside the disk. Whereas a magma layer inside the Roche limit is not inconsistent with our findings (the moon or moons that would have formed from this layer would have impacted Mars a long time ago), future modeling of the disk should account for an extended disk where the pressure and the temperature allow for a direct condensation of the vapor into solid grains.

In order to provide constraints for future models of the martian disk, we determined the pressure-temperature ranges where the gas would directly condense into solid grains assuming a BSM composition of the gas. We restricted our analysis to the condensation of olivine only. 
\citet{gai98} found that olivine would first condense as nearly pure forsterite and iron might be included later on in the solution. We consequently considered the condensation of pure forsterite. The pressure up to which solid or liquid forsterite is stable for a given temperature can be calculated with the following formula:
\begin{equation}
  P(T)^3=\frac{1}{x^2_{\mathrm{MgO}}x_\mathrm{SiO_2} e^{-\Delta G_{s,l}/RT}}
\end{equation}
where $\Delta G_{s,l}$ is the Gibbs free energy of formation of solid or liquid forsterite from gaseous MgO and $\mathrm{SiO_2}$ (it can be calculated with the JANAF online tables), and $x_{\mathrm{MgO}}$ and $x_\mathrm{SiO_2}$ are the molar fraction of the gases. An extensive chemical model would thus be required in order to infer these molar fraction. As such detailed modeling is beyond the scope of the present work, we simply assumed that different fractions ($f$) of Mg and Si were bound into MgO and $\mathrm{SiO_2}$ molecules in the gas phase and thus $x_\mathrm{MgO,SiO_2}=f\epsilon_\mathrm{Mg,Si}$, $\epsilon$ being the fraction of the element.

\begin{figure*}
\begin{center}
\includegraphics[scale=0.6]{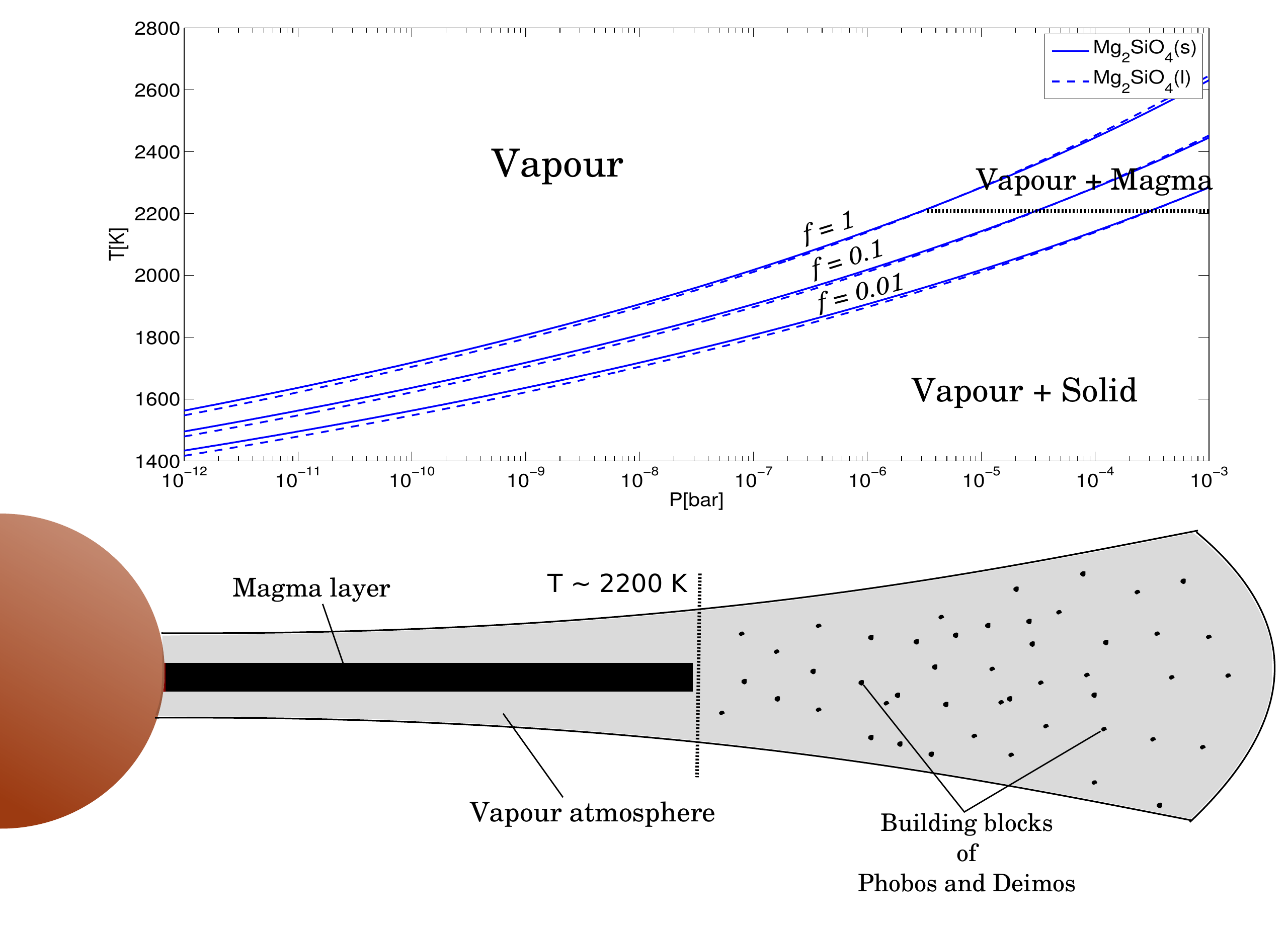}
\caption{Schematic representation of the plausible structure of the accretion disk around Mars. The external part of the disk where Phobos and Deimos may have formed is defined as the region on the right side of the vertical line where the temperature drops below $\sim$2200 K and solids start to condense. The location of the synchronous orbit shall lie in this extended part of the disk to prevent rapid orbital decay of the moons towards Mars.The equilibrium curves of solid (solid line) and liquid (dotted line) forsterite are shown for different fractions ($f$) of Mg and Si bound into MgO and $\mathrm{SiO_2}$. }
\label{fig2}
\end{center}
\end{figure*}

The stability curves are shown in Figure~\ref{fig2} for $f=1$, $f=0.1$ and $f=0.01$. The last two cases are more realistic given that Mg is usually found as a free atom whereas Si is mainly found in SiO molecules. We find that the solid  phase of forsterite becomes more stable than the liquid one below a temperature of $\sim 2200 \, \mathrm{K}$. At this temperature, the condensation of forsterite occurs at a pressure lower than $[10^{-6}-10^{-3}] \, \mathrm{bar}$ depending on the partial pressures of MgO and $\mathrm{SiO_2}$. 


\section{Discussion}

Here, we have opened the possibility that gas-to-solid condensation in the external part of an extended gaseous disk is a likely formation mechanism for the martian moons building blocks as it would lead to the formation of small ($\leq$ 2 microns) dust particles. Accretion from such tiny grains would naturally explain the similarity in spectral properties between D-type asteroids (or comets) and the martian moons as well as their low densities. It therefore appears that accretion in the external part of an impact-generated gaseous disk is a likely formation mechanism for the martian moons that would allow reconciling their orbital and physical properties. Future work should attempt modeling the mineralogy resulting from the gas-to-solid condensation sequence and verify that a space weathered version of the derived composition sieved to sub-micron sized grains is compatible with the spectral properties of the moons. Such investigation would greatly benefit form detailed numerical models that would constrain the thermodynamic properties of a circum-Mars impact generated disk as a function of radial distance. 

Finally, note that our proposed scenario is not incompatible with the the presence of weak hydration features in the spectra of Phobos and Deimos \citep{giu11,fra12,fra14}. Water has been delivered to the surfaces of most if not all bodies of the inner solar system, including the Moon \citep{sun09}, Mercury \citep{law13} and Vesta \citep{scu15}. It has thus become clear over the recent years, that space weathering processes operating at the surfaces of atmosphere-less inner solar system bodies do not only comprise the impact of solar wind ions and micrometeorites, which tend to redden and darken the spectra of silicate-rich surfaces, but they also comprise contamination and mixing with foreign materials including water-rich ones \citep{pie14}.

\acknowledgments
We warmly thank the referee for his very constructive review.  P.V. and O.M. acknowledge support from CNES. This work has been partly carried out thanks to the support of the A*MIDEX project (n\textsuperscript{o} ANR-11-IDEX-0001-02) funded by the ``Investissements d'Avenir'' French Government program, managed by the French National Research Agency (ANR).

\section{APPENDIX}

\subsection{Grain size resulting from magma solidification}

The grain size of a crystal resulting from magma solidification has been extensively studied and appears closely related to its cooling history. Rapid cooling ( $\gtrsim 10^{2}$ $\mathrm{K\, hr^{-1}}$) will lead to the formation of smaller grains ($\sim 10^{-2}$ mm) whereas slow cooling ( $\lesssim 1$ $\mathrm{K\,hr^{-1}}$) will lead to the formation of larger grains ($\sim 1$ mm) as is observed in experimental studies and Earth's samples \citep{fle76,Ich85,gro90,cas88,cas93}. This is well illustrated in the case of Earth's rocks via basalts and gabbros. These rocks possess the same composition but a very different texture. Basalts are extrusive igneous rocks that experienced rapid cooling either within Earth's atmosphere or oceans and possess a fine-grained structure. On the other end, gabbros are intrusive igneous rocks that crystallized below Earth's surface on much longer timescales and thus exhibit coarse grains. 

In the present case, assuming that the temperature of the magma layer is regulated by the disk radiative cooling only, an order of magnitude of the disk cooling rate is :
\begin{equation}
  \frac{dT}{dt} = -\frac{ 2 \pi R^2_\mathrm{disk} \sigma_{SB} T^4_{ph} }{ M_\mathrm{disk} C_p} \sim - 2 \; \mathrm{K\,hr^{-1}}
\end{equation}
where $R_\mathrm{disk} = 4 R_\mathrm{Mars}$ which is approximately the Roche limit, $\sigma_{SB} = 5.67 \times 10^{-8} \; \mathrm{J\,m^{-2}\,K^{-1}}$ is the Stefan-Boltzmann constant, $T_{ph} = 2000$ K is the temperature of the disk at the photosphere that is maintained at $\sim 2000 \;\mathrm{K}$ throughout its lifetime \citep{war12}, $M_\mathrm{disk} = 5\times 10^{20} \; \mathrm{kg}$ following the results of \citet{cit15} and $C_p = 4 \times 10^{3} \; \mathrm{J\, kg^{-1} \, K^{-1}}$ is the heat capacity of the vapor. 

One may also consider that the clumps formed still molten at the Roche limit. In this case, assuming a density $\rho = 3300 \; \mathrm{kg\,m^{-3}}$ for the melt and clumps with typical sizes in the 1-10 km range, an order of magnitude of the clump cooling rate would be :
\begin{equation}
  \frac{dT}{dt} = - \frac{3 \pi \sigma_{SB} T^4_s}{R_\mathrm{clump} \rho C_p} \sim 0.2\!-\!2 \; \mathrm{K\,hr^{-1}}
\end{equation}
where the largest clumps would possess the slowest cooling rates and vice versa.  In either case, the cooling timescales are comparable to those derived from laboratory experiments and the rocks that would have crystallized from the magma should typically exhibit the same grain sizes as those found in magmatic rocks on Earth. 

In a different register, it is interesting to note that chondrules, which formed as molten or partially molten droplets in space before being accreted to their parent asteroids, have typical sizes in the 0.1-1 mm range \citep{Hut04}. In summary, magma condensates appear always coarse grained (grain size in the 0.1-1mm range), regardless of their formation mechanism.



\subsection{Grain size resulting from gas to solid condensation}

The texture and size of the grains condensing directly from the vapor is, similarly to the solidification from a magma, closely related to the cooling rate of the vapor. 
Fast cooling will be associated with a high nucleation rate. As such, a rapid decrease of the temperature will imply the condensation of a large number of small grains. On the contrary, if the cooling rate is slow, fewer nuclei will condense but these will grow by continuous condensation of vapor onto their surface which will result, on average, in the development of larger grains. This process has been investigated via both theory (\citet{gai84} and \citet{gai88}), and laboratory experiments (\citet{rie99a}, \citet{rie99b}, \citet{rie99c}, \citet{top06}, \citet{des16}). The theoretical investigations of grain growth in stellar outflows was applied to carbon growth and resulted in grains with sizes in the $10^{-3}$--1 $\mu$m range \citep{gai84}.
Concerning the laboratory experiments, two cases have been investigated namely a rapid and a slow condensation of silicate rich vapor into solid grains. In the case of rapid condensation which occurs in non-equilibrium, the formation of very small amorphous grains with typical sizes of a few tens to a few hundred nanometers was observed, in agreement with theoretical calculations by \citet{gai84}. In the case of slow condensation which occurs near equilibrium, the formation of small crystalline grains with typical sizes of a few hundred nanometers was observed \citep{top06}.
It therefore appears that - in either case (slow and fast cooling) - small dust grains with typical sizes of $\sim$0.1 microns are the natural outcome of gas to solid condensation. This is very coherent with the typical grain sizes observed among interplanetary dust particles \citep{rie09}.

\clearpage

\begin{deluxetable}{lcccc}
\tablecaption{Bulk Silicate Compositions used to model the disk composition}
\tablehead{
\colhead{Oxide Wt\%} 	& \colhead{BSM$^a$} 	& \colhead{dep. BSM$^b$} 	& \colhead{Moon$^c$} 	& \colhead{IDP$^d$}}
\startdata
$\mathrm{SiO_2}$ 		& 45.39 			& 17.4 				& 44.60 			& 47.00  \\
MgO 				& 29.71 			& 20.5 				& 35.10 			& 16.8 \\
MnO 				& $-$  			&  $-$  				& $-$ 			& 0.1 \\
NiO 					& $-$ 			& $-$ 				& $-$ 			& 1.1 \\
$\mathrm{Al_2O_3}$		& 2.89 			& 2.19 				& 3.90 			& 1.3 \\
$\mathrm{TiO_2}$ 		& 0.14 			& 0.09 				& 0.17 			& $-$ \\
Feo 					& 17.21 			& 10.55 				& 12.40 			& 24.4 \\
CaO 					& 2.36 			& 1.81 				& 3.30 			& 0.9 \\
$\mathrm{Cr_2O_3}$ 	& $-$ 			& $-$	 			& $-$ 			& 0.2 \\
$\mathrm{Na_2O}$ 		& 0.98 			& 0.01 				& 0.05 			& $-$ \\
$\mathrm{K_2O}$ 		& 0.11 			& 0.01 				& 0.004 			& $-$ \\
S 					& $-$ 			& $-$ 				& $-$ 			& 7.3 \\
\\
Total 				& 98.79 			& 52.54 				& 99.5 			& 99.09 \\
\enddata
\tablecomments{$^a$Bulk Silicate Mars \citep{lod11}; $^b$depleted BSM estimated for a 50\% vapourized disk \citep{can15}; $^c$Bulk Silicate Moon \citep{one91}; $^d$Interplanetary Dust Particle \citep{rie09}.}
\label{tbl1}
\end{deluxetable}

\clearpage
\begin{deluxetable}{lcccccc}
\tabletypesize{0.7\scriptsize}
\tablecaption{Bulk Mineral Composition resulting from the CIPW norm calculation \label{tbl2}}
\tablehead{
\colhead{Minerals Wt\%} & \colhead{BSM} & \colhead{dep. BSM} & \colhead{Moon} & \colhead{Moon/BSM} & \colhead{IDP} & \colhead{IDP/BSM}\\
{} & {100\%} & {100\%} & {100\%} & {50\%/50\%} & {100\%} & {50\%/50\%}
}
\startdata
Quartz & $-$ & $-$ & $-$ & $-$ & 9.83 & $-$ \\
Plagioclase & 11.59 & 10.45 & 10.88 & 11.24 & 3.62 & 9.57 \\
Orthoclase & 0.66 & $-$ & 0.02 & 0.34 & $-$ & 0.33 \\
Diopside & 6.97 & $-$ & 4.78 & 5.87 & 0.76 & 5.69 \\
Hypersthene & 21.29 & $-$ & 23.24 & 22.24 & 66.00 & 48.86 \\
Olivine & 59.22 & 84.65 & 60.76 & 60.01 & $-$ & 27.58 \\
Magnetite & $-$ & 2.97 & $-$ & $-$ & 4.01 & $-$ \\
Pyrite & $-$ & $-$ & $-$ & $-$ & 15.78 & 7.84 \\
\\
Total & 99.73 & 98.07 & 99.68 & 99.70 & 100 & 99.87  \\
\enddata
\end{deluxetable}

\end{document}